\documentclass{tCPH2e}

\usepackage{epstopdf}% To incorporate .eps illustrations using PDFLaTeX, etc.
\usepackage{subfigure}% Support for small, `sub' figures and tables

\theoremstyle{plain}
\newtheorem{theorem}{Theorem}[section]
\newtheorem{corollary}[theorem]{Corollary}
\newtheorem{lemma}[theorem]{Lemma}

\theoremstyle{definition}
\newtheorem{definition}{Definition}

\theoremstyle{remark}

\begin{document}

%\jvol{00} \jnum{00} \jyear{2015} \jmonth{January}

%\articletype{GUIDE}

\title{Revisiting Interval Graphs for Network Science}

\author{
\name{Chuan Wen, Loe\textsuperscript{a}$^{\ast}$\thanks{$^\ast$Corresponding author. Email: c.loe11@imperial.ac.uk}
and Henrik Jeldtoft Jensen\textsuperscript{a}}
\affil{\textsuperscript{a}Department of Mathematics and Centre for Complexity Science, \\Imperial College London, London, SW7 2AZ, UK}
\received{v4.0 released January 2015}
}

\maketitle

\begin{abstract}
The vertices of an interval graph represent intervals over a real line where overlapping intervals denote that their corresponding vertices are adjacent. This implies that the vertices are measurable by a metric and there exists a linear structure in the system. The generalization is an embedding of a graph onto a multi-dimensional Euclidean space and it was used by scientists to study the multi-relational complexity of ecology. However the research went out of fashion in the 1980s and was not revisited when Network Science recently expressed interests with multi-relational networks known as multiplexes. This paper studies interval graphs from the perspective of Network Science.
\end{abstract}


%{\abstractfont\centerline{\bfseries Index to information contained in this guide}\vspace{12pt}
%\hbox to \textwidth{\hsize\textwidth\vbox{\hsize19pc
%\hspace*{-12pt} {1.}    Introduction\\
%{2.}    Preliminaries\\
%\hspace*{10pt}{2.1.}  Interval Graphs\\
%\hspace*{10pt}{2.2.}  Interval Graphs as Hyper-Boxes\\
%\hspace*{10pt}{2.3.}  Topology of an Unknown Structure\\
%\hspace*{24pt}            (Example)\\
%\hspace*{10pt}{2.4.}  Multiplexes\\
%{3.}    Structural Connection Between Interval\\
%\hspace*{8pt}            Graphs and Multiplexes\\
%\hspace*{10pt}{3.1.}  The Union of a Multiplex\\
%\hspace*{10pt}{3.2.}  The Intersection of a Multiplex\\
%{4.}    Information Propagation of Interval \\
%\hspace*{8pt}            Graphs\\
%\hspace*{10pt}{4.1.}  Linear Fine Structures\\
%\hspace*{24pt}{4.1.1}  Overview\\
%\hspace*{24pt}{4.1.2}  Evolutionary Interval Graphs\\
%\hspace*{24pt}{4.1.3}  Phase Transition\\
%\hspace*{10pt}{4.2.}  Propagation Models\\
%\hspace*{24pt}{4.2.1}  Infection Model\\
%\hspace*{24pt}{4.2.2}  Influence Model\\
%\hspace*{10pt}{4.3.}  Experimental Results\\
%{5.}    Approximating Boxicity Using\\ 
%\hspace*{8pt}            Communities Detection \\
%\hspace*{10pt}{5.1.}  Minimum Boxicity of Network from\\
%\hspace*{24pt}            its Communities\\
%\hspace*{10pt}{5.2.}  Boxicity of the Communities'\\ 
%\hspace*{24pt}            Interaction Network\\
%\hspace*{10pt}{5.3.}  Boxicity with Experimental Noise\\
%{6.}    Discussions\\}}}

\section{Introduction}
The vertices of an interval graph represent intervals over a real line where overlapping intervals denote that their corresponding vertices are adjacent. This implies that the vertices are measurable by a metric and there exists a linear structure in the system. For example interval graphs was introduced to deduce the linearity of genes when Benzer noticed that the behavior of mutated strains of bacteriophage T4 (virus) forms an interval graph \cite{benzer1959topology}.

The generalization of interval graphs is where the vertices are $d$-dimensional axis-parallel  hyper-boxes such that intersecting boxes implies that their corresponding vertices are adjacent in the graph $G$. This can be expressed as a finite set of $d$ interval graphs on the same vertex set, i.e $\mathcal{B}=\{I^1,\ldots,I^d\}$ such that $G=(V,E_1 \cap\ldots\cap E_d)$ where the interval graph $I^k(V,E_k)$ is the projection of the boxes onto the $k^{th}$ axis. 

This can be visualized by taking the species in an ecology as boxes and each of the axes measures a different environmental factor like temperature, soil acidity, amount of sunlight, etc. Each species are enclosed in their unique environment phase space where they are adaptable, and the boxes that intersect implies that the species can coexist in a common environment.

Thus interval graphs are used to model the stability and complexity of ecology system by studying the number of factors (dimensions) the species in the ecology depend on \cite{opac-b1129706,eklof2013dimensionality,jordan2004network}. Applications of interval graphs also arise naturally in many time dependence problems like task scheduling \cite{carlisle1995k,kolen2007interval} or other linear structures like pavement deterioration analysis \cite{gattass1981} and Bioinformatics \cite{jungck1982computer,biedl2004finding}.

After 20 years of research, interval graph went out of fashion in the 1980s and was not revisited when Network Science recently expressed interests on multi-relational networks like  \emph{multiplex} \cite{Kivela2013,boccaletti2014structure}. Multiplexes and interval graphs belong to the same mathematical object known as \emph{graphs on the same vertex set}, where a multiplex is a collection of graphs $\mathcal{M}=\{G^1(V,E_1),\ldots,G^d(V,E_2)\}$ with each graph representing the different relationships of a system.

For example in a social multiplex, people are connected by specific relationship categories like friends, colleagues or family. Other examples include the different modes of transportations in a transport system and the different ways researchers are connected in citation networks \cite{Kivela2013}. It is the modern outlook of Network Science to preserve the rich relational data of the system. 

Specifically interval graphs and multiplexes are special types of a more general mathematical object known as \emph{intersection graph} \cite{terry1999topics}, where a vertex pair is connected if they have overlapping attributes. E.g. in a social multiplex, two individuals are connect as schoolmates (relationship) if they are in the same school together (attribute).

The motivation of this paper is to show that the tools/perspectives from Network Science can benefit interval graphs research, vice versa.

\section{Preliminaries}
\subsection{Interval Graphs}
\begin{figure}
     \centering
     \includegraphics[width=0.6\textwidth]{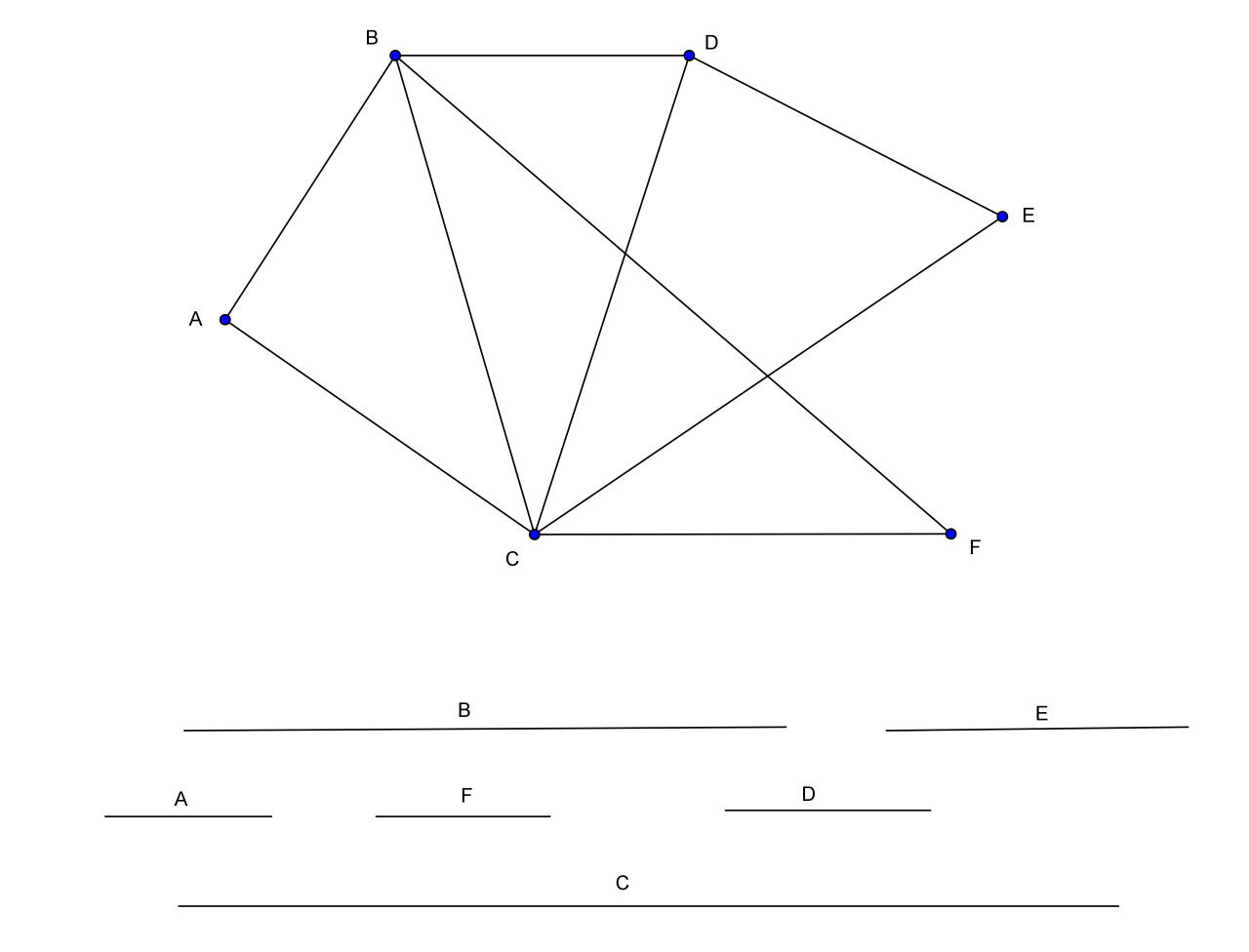}
     \caption{The duality of an interval graph (above) and a set of intervals 
     (below). There is a bijective map between the vertices of the graph and 
     the intervals where overlapping intervals denote the adjacency of their 
     corresponding vertices. For example interval $A$ overlaps interval $B$ 
     implies that vertex $A$ is adjacent to vertex $B$, vice versa.}
     \label{fig:interval_graph}
\end{figure}

\begin{definition}
\label{def:interval}
An interval graph $I(V,E)$ maps a set of intervals $\{J^1,\ldots,J^n\}$ as vertices such that adjacent vertices $(a,b)$ denotes $J^a\cap J^b \neq \emptyset$ \cite{fishburn1985interval} (Fig. \ref{fig:interval_graph}). 
\end{definition}

The sequential nature of the intervals implies that there is a linear ordering $\prec$ on the vertices where for all vertex triples $v_1,v_2,v_3 \in V$ with $v_1 \prec v_2$ and $v_2 \prec v_3$, if $(v_1,v_3) \in E$ then by transitivity $(v_1,v_2), (v_2,v_3) \in E$. This colloquially states that there is no ``shortcut" in the graph, i.e. there is no independent vertex triples where every two of them are connected by a path avoiding all neighbors of the third. This property is known as \emph{asteroid-triple free} (AT-free). 
\begin{theorem}\label{interval_test}
An interval graph is chordal and AT-free \cite{Lekkeikerker1962}.
\end{theorem}

The lack of ``shortcut" in AT-free graphs restricts the number of paths among the vertices in the graph and hence limits the search space for a variety of problems. Thus the AT-free property presents useful algorithmic structure on interval graphs such that some NP-complete graph problems are tractable in polynomial time \cite{chandran2010geometric}.

\subsection{Interval Graphs As Hyper-Boxes}\label{sec:hyperbox}
\begin{figure}
     \centering
     \includegraphics[width=1.0\textwidth]{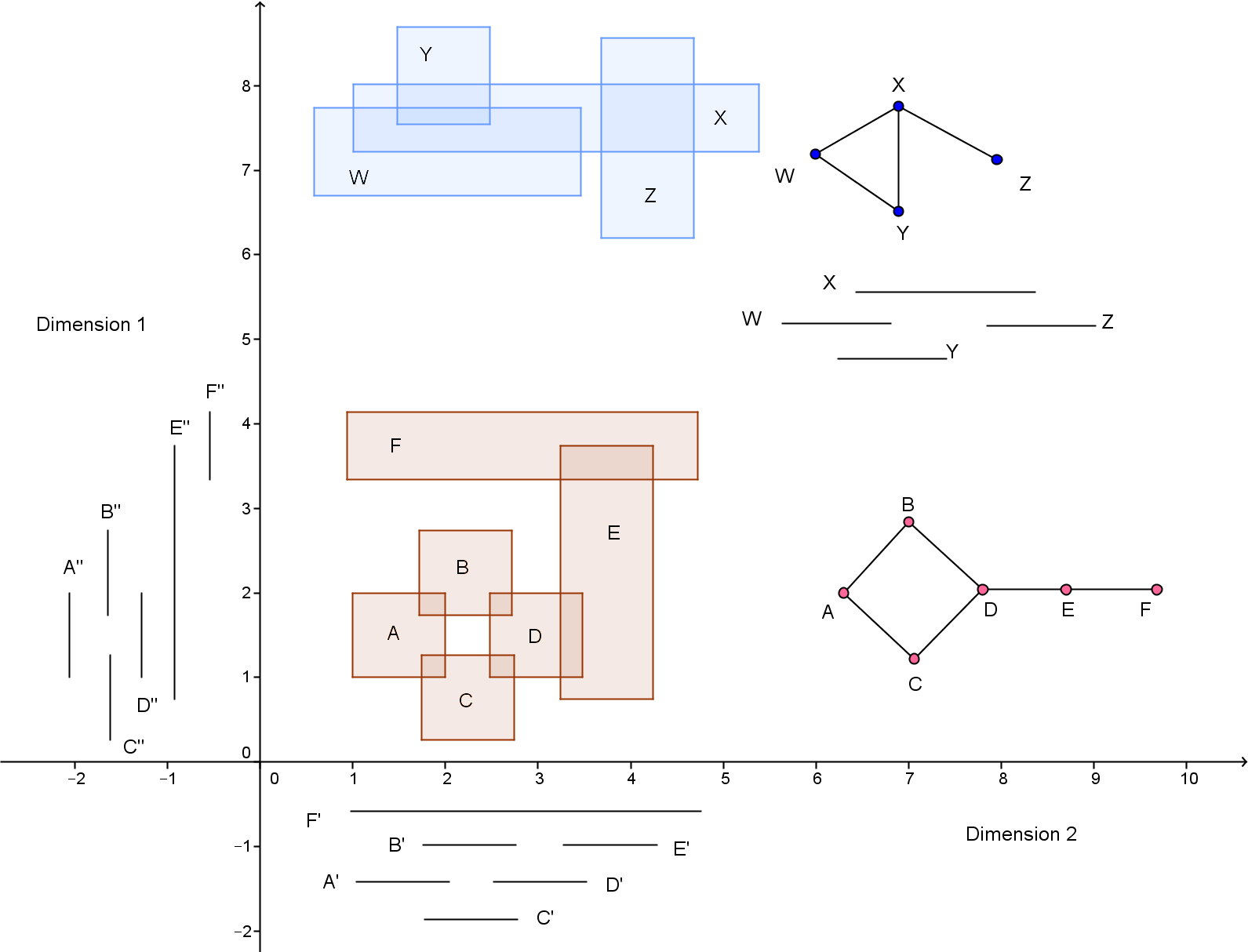}
     \caption{The set of 2-dimensional boxes $A,\ldots,F$ corresponds to the 
	     graph on its right, and they are from the intersection of 2 interval 
	     graphs with vertex labels $A',\ldots,F'$ (dimensional 1) and 
	     $A'',\ldots,F''$ (dimensional 2). Adjacent vertices are 
	     equivalent to saying that their respective boxes intersect. 
	     However a graph constructed from $m$-dimensional boxes does not 
	     necessary implies that it has boxicity $m$. For instance the 
	     top graph with vertex labels $X,\ldots,Z$ is constructed with  
	     $m=2$-dimensional boxes, but since it can be represented with 
	     a 1-dimensional interval graph, its boxicity is one.}
     \label{fig:interval}
\end{figure}

The graph from the intersection of interval graphs $I^i(V,E_i)$, i.e. $G(V,E_1\cap\ldots\cap E_m)$ forms a set of axis-parallel hyper-boxes as vertices in $m$ dimensions, and the adjacent vertices implies that their corresponding hyper-boxes intersects. The minimum $m$ interval graphs to represent $G$ is its \emph{boxicity} and it is a measure of complexity (Fig. \ref{fig:interval}) \cite{opac-b1129706,stouffer2006robust}.

For instance the food webs in ecology is a \emph{competition graph}, where two species (vertices) are connected if they compete over the same food source. Although ecology is generally known to be a complex system, Cohen showed that food webs are generally low dimensional and simple. In fact many food webs are interval graphs where the ordering of the intervals (as predators) correlates to the size of their preys \cite{opac-b1129706}.

A marine food web is an example of a food web that is not an interval graph as a predator feeds on species based on two environment niches --- the size of the prey and the depth of the water. Since each of the niches can be expressed as an interval graph, two predators are in competition if they feed on the same preys, i.e. their 2-dimensional boxes intersect.

However it is computationally hard (NP-complete) to determine the boxicity of an arbitrary graph as there are more degrees of freedom to embed a graph in $d$-dimensional space. Fortunately one can still make approximations by the analytical bounds on the degree of the graph. Suppose $\Delta$ and $\delta$ are the maximum and minimum degree of a $n$-vertices graph respectively, the boxicity of the graph is bounded between $n/(2(n-\delta-1))$ 
\cite{adiga2008lower} and $min(n/2,\Delta^2+2,\lceil(\Delta+2)\ln{n}\rceil)$ \cite{sunil2008boxicity}.

The boxes are synonymous to embedding its graph in $m$ dimensional Minkowski r-metric space $M^r_{m}$ such that for all adjacent $u,v \in V$, their distance in the metric space is bounded by some length \cite{Freeman83}:
\begin{equation}\label{metric_dist}
d_{uv}(\langle f_1(u),\ldots,f_{m}(u)\rangle , 
       \langle f_1(v),\ldots,f_{m}(v)\rangle) \leq l_u + l_v, 
\end{equation}
where $l_u$ and $l_u$ are length given for their respective vertices, and $\langle f_1(u),\ldots,f_{m}(u)\rangle$ is a vector mapping  $u$ to the metric space with the real-value functions $f_1,\ldots,f_{m}$. In addition the functions $f_i$ on all $u,v \in V$ is conditioned by Minkowski r-metric space:
\begin{equation}
d_{uv} = \Big[ \sum^{m}_{i=1} |f_i(u) - f_i(v)|^r \Big]^{1/r}. 
\end{equation}

The arbitrary constant $r$ is a weighting parameter where all components $|f_i(u) - f_i(v)|$ are equally weighted for $r=1$ (i.e. Manhattan Distance). For $r=2$ (i.e. Euclidean Distance), the components that are greater contribute more to the Minkowski Distance. Hence by letting $r=\infty$ to complete the metric space, the greatest component will dominate the distance where $d_{uv}=\max_{i=1}^{m}|f_i(u)-f_i(v)|$, and each point is a hyper-box with sides parallel to the axes.

\subsection{Topology of an Unknown Structure (Example)}
\label{sec:determine_topology}
\begin{figure}
     \centering
     \includegraphics[width=0.9\textwidth]{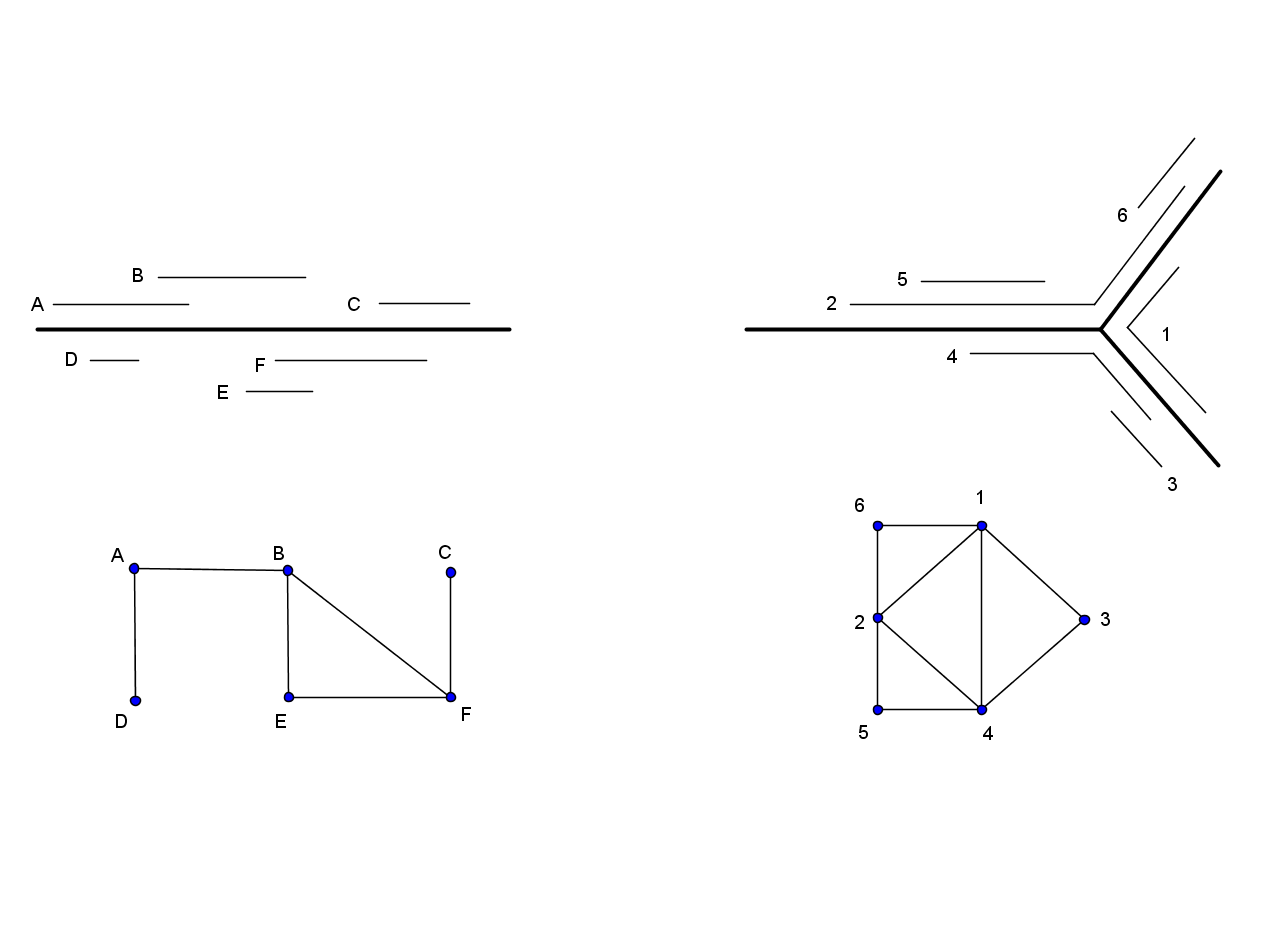}
     \caption{A comparison of a linear structure (left) and a branched 
		structure (right). Adjacent vertices denote that their respective segments 
		overlap (e.g. vertex $C$ is adjacent with vertex $F$ as segment $C$ 
		overlaps with segment $F$). Since the graph on the left corresponds to a 
		linear structure, it is an interval graph. The graph on the right is not an 
		interval graph as vertices 3, 5 and 6 form an asteroid-triple --- path 
		3-1-6 (3-4-5 and 5-2-6) avoids the neighbors of vertex 5 (respectively 6 
		and 3).}
     \label{fig:gene}
\end{figure}

Benzer deduced the linearity of genes by rejecting the hypothesis that it is from a non-linear structure with interval graphs. Suppose there are two hypotheses of a gene's structure --- linear and branched (Fig. \ref{fig:gene}). 

The vertices are the different mutated variants of the T4 virus such that they do not have the complete genome to kill bacterias independently. In this case vertex $A$ refers to the variant where segment $A$ of T4 is changed. If viruses with overlapping segments do not have the entire information to kill the bacterias, then an edge is placed between them. Since the graph on the left is constructed from a linear structure, it is an interval graph. 

However if T4 genes was a branched structure, then the resultant graph will not be an interval graph. In the same figure, vertices 3, 5 and 6 form an asteroid-triple --- the path 3-1-6 (3-4-5 and 5-2-6) avoids the neighbors of vertex 5 (respectively 6 and 3). It is noteworthy to observe that by removing any of the vertices 1, 2 or 4, the graph will be an interval graph. It is also possible to get an interval graph by removing edges $\{1,3\}$ and $\{1,4\}$ from the original graph. Thus interval graphs only \emph{supports} the hypothesis of a linear structure, but it is insufficient to \emph{prove} the linearity of a system.

\subsection{Multiplex}
\begin{definition}\label{def:multiplex}
A multiplex is a finite set of networks, $\mathcal{M} = \{G^1,\ldots,G^m\}$, where every graph $G^i(V,E_i)$ has a distinct edge set $E_i \subseteq V \times V$. 
\end{definition}

Multiplex is a natural transition from network as a model to preserve the rich relational properties in the data. Each of the networks refers to a distinct relationship in the complex system, e.g. a transportation complex system has different modes of transport and each type can be represented by one of the networks in the multiplex. 

However the relationships in many multiplexes are not well defined by physical infrastructures like the transportation complex systems. For instance the relationships in a social multiplex like colleagues, family, friends, etc are chosen based on the researchers' opinions or by the limitation of their data. Thus a system can be easily expressed as two distinct multiplexes and dynamics. As a result the reliability from the conclusions of such multiplexes can be unstable.

Therefore one of the goal of this paper is to introduce interval graphs as a way to study the granularity of such relationships. If we assume that a linear structure like an interval graph is the simplest relational behavior of a complex system, then by using the concept of boxicity, the \emph{guideline} is that the number of relationships of a multiplex should not be more than the boxicity of the multiplex's projection (details in the next section).

\section{Structural Connection Between Interval Graphs and Multiplexes}
\subsection{The Union of a Multiplex}\label{sec:union_multiplex}
\begin{definition}\label{def:projection}
The projection of a multiplex $\mathcal{M} = \{G^1(V,E_1),\ldots,G^m(V,E_m)\}$ is the graph from the union of all the edge sets, i.e. $\bar{G}(V,E_1\cup\ldots\cup E_m)$.
\end{definition}

Before the popularization of multiplex research, many scientists tend to simplify their multi-variate data by projection. For example the Zachary Karate Club Network is the projection of 8 networks (relationships) on the same vertex set \cite{zachary1977ifm}. It is the most general way to describe the connectivity of a system.

In the previous section we suggested that the number of relationships in a multiplex should not exceed the boxicity of its projection. Given that it is possible to express the same (projected) network with a simpler model like low dimensional hyperboxes, a multiplex with a higher number of relationships will appear unnecessarily complex. Thus unless there are justifications for a high number of relationships, deviating this guideline is synonymous to going against the grain of Occam's Razor Principle.

Another argument for this guideline is to assume that any dynamics of a system are measurable by some metric. For example the acquaintanceship of a school alumni social network can be ``measured" by the time when the members attended the school. Friends within the alumni are often schoolmates at around the same period. 

However some dynamics are more complex and require more than one metric to measure them, for instance the dynamics of prey-and-predator in a marine ecology (section \ref{sec:hyperbox}). Thus we posit that if every dynamics is measurable by at least one (linear) metric, then the number of metrics is the upper bound to the number of dynamics. Hence a multiplex with significantly more relationships than the boxicity of its projection suggests that some of the relationships are driven by the same dynamics and it maybe more appropriate to combine these relationships.

The above arguments are not rigorous and hence we emphasize that it should be taken as a guideline. However the challenge to this approach is that to determine the boxicity of an arbitrary network is a NP-complete problem and could be the same reason to why boxicity is not common in Complexity Science. The degree distribution and clustering coefficient of the union of networks ensembles can be found in \cite{loe13}.

\subsection{The Intersection of a Multiplex}\label{sec:intersection_multiplex}
\begin{definition}\label{def:intersect}
The intersection of a multiplex $\mathcal{M} = \{G^1(V,E_1),\ldots,G^m(V,E_m)\}$ is the graph from the intersection of all the edge sets, i.e. $H(V,E_1\cap\ldots\cap E_m)$.
\end{definition}

The projection of a multiplex appears to be the counter-thesis of modern Network Science by reducing the problem back to a network. However in order to understand the connection between multiplexes and networks, it is important to study the process from both sides. Hence without compromising too much relational information for simplicity, the analysis of the overlapping edges from the projection is pivotal, i.e. the graph $H(V,E_1\cap\ldots\cap E_m)$.

\subsubsection{Statistical Structural Properties}
The distribution of the overlapping edges is an essential characteristic to distinguish multiplex ensembles from random \cite{Bianconi13,chen2012degree,Lee12,RanolaASSL10}. For example a multiplex ensemble is defined to be correlated if the expected number of overlapping edges deviates from the behavior of a collection of random Erd\H{o}s-R\'{e}nyi graphs \cite{Bianconi13}. The degree of correlation in turn affects the phase transition of the emergence of a giant component \cite{Lee12}.

However besides the trivial case on the intersection of Erd\H{o}s-R\'{e}nyi networks \cite{loe13}, we have little understanding on the statistical properties like degree distribution or clustering coefficient of the intersection of the different network ensembles. In this paper, we will present the intersection of two combinations --- Erd\H{o}s-R\'{e}nyi network with Barab\'{a}si-Albert network, and Watts-Strogatz network with Barab\'{a}si-Albert network.

\vspace{10pt}
\noindent
\textbf{Erd\H{o}s-R\'{e}nyi network with Barab\'{a}si-Albert}: Barab\'{a}si-Albert graph $B_{n,m}$ on $n$ vertices is an error-free model to simulate the preferential attachment phenomenon in real world network, where $m$ new edges are added at each iteration \cite{barabasia99}. Therefore to make it more relevant to real-world problems, this has lead to Barab\'{a}si-Albert variants with experimental noise \cite{dorogovtsev2000structure,Pennock02winnersdont} in which preferential and random uniform (noise) attachment are combined. Similarly in the projection of a multiplex with Erd\H{o}s-R\'{e}nyi and  Barab\'{a}si-Albert graph, the former adds the uniform attachment noise to the scale-free system \cite{loe13}. 

Let $p$ be the probability that a pair of vertices are connected in the Erd\H{o}s-R\'{e}nyi network. When the Erd\H{o}s-R\'{e}nyi network intersects a Barab\'{a}si-Albert network, only a fraction (i.e. $p$) of the edges that is incident to vertex $v$ are in the edge set of both networks. Hence if $v$ has degree $k$, then after the intersection its resultant degree is $\approx kp$. 

Define $P_r$ and $P_b$ to be the degree distribution of the resultant network and Barab\'{a}si-Albert respectively. Suppose we want to find $P_r(deg = x)$, we can group all the vertices in Barab\'{a}si-Albert that are most likely to have degree $x$ after the intersection, i.e. $\lfloor kp \rfloor = x$. Thus: 
\begin{equation}
P_r(deg = \lfloor kp \rfloor) \approx \sum_{i=0}^{\lfloor 1/p \rfloor} P_b(deg = \lfloor kp \rfloor + i), 
\end{equation}
or 
\begin{equation}\label{eq:ERBA}
P_r(deg = x) \approx \sum_{i=0}^{\lfloor 1/p \rfloor} P_b(deg = \lceil x/p \rceil + i), 
\end{equation}
where $P_b(deg = k) \sim k^{-3}$ is scale-free.

Note that $P_b(deg = \lceil x/p \rceil + i) \sim \lceil x/p \rceil^{-3}$ as $x/p$ dominates all values of $i$. Hence $P_r(deg = x) \approx \lfloor 1/p \rfloor \cdot \lceil x/p \rceil^{-3}$, implying that the subgraph is also scale-free. However this does not contradict the conclusion of  \cite{Stumpf05} where ``the subgraphs of scale-free networks are not scale-free". What Stumpf had done was to derive a subgraph by removing vertices of a scale-free network whereas in our case it is only the edges of a scale-free network that are removed.

\vspace{10pt}
\noindent
\textbf{Watts-Strogatz network with Barab\'{a}si-Albert network}: A real-world characteristic that Barab\'{a}si-Albert ensemble fails to model is the likelihood that vertices tend to cluster together in graphs, i.e. high clustering coefficient. This characteristic can be modeled with a  
Watts-Strogatz $W_{n,w,q}$ on $n$ vertices where $w$ is the degree of a ring lattice for the initial construction, and $q$ is the rewiring probability.

\begin{figure}
  \centering
  \includegraphics[width=1.0\textwidth]{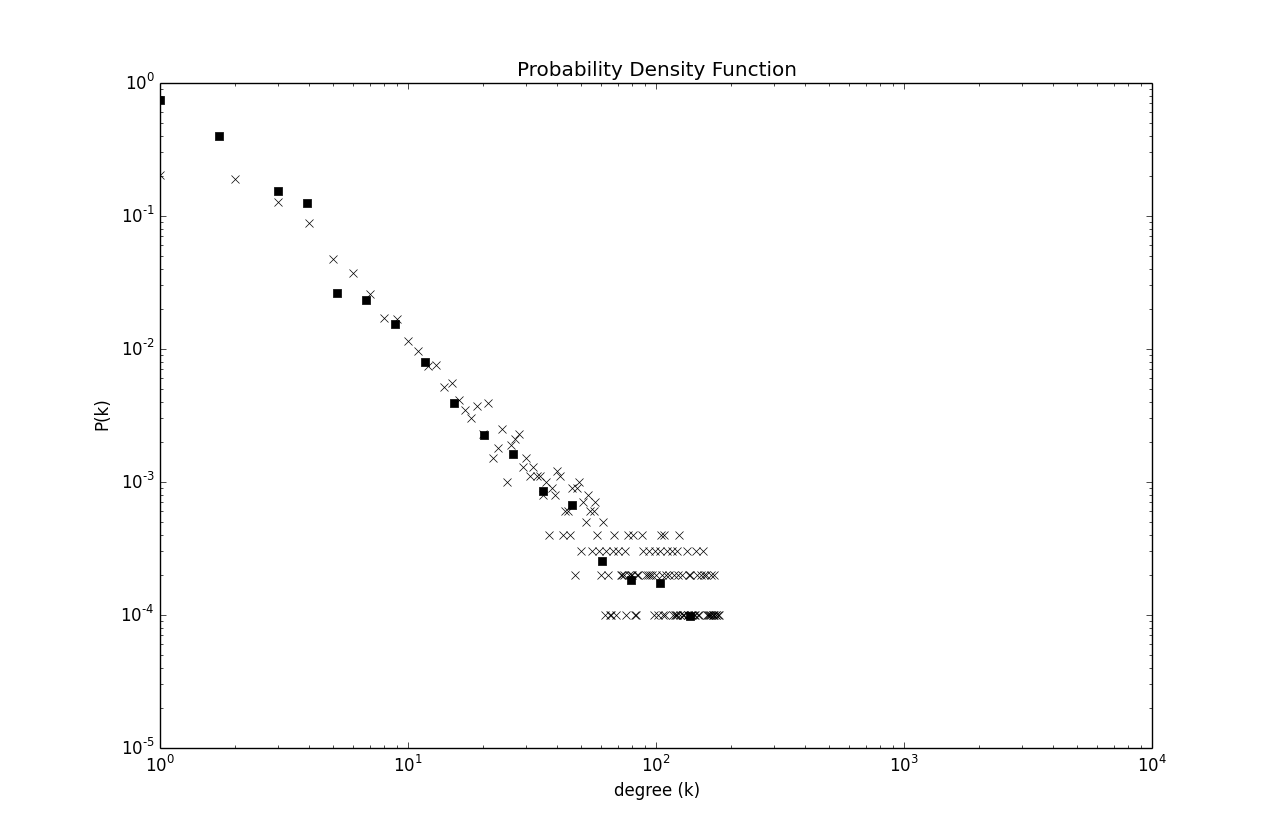}
    \caption{The intersection of Watts-Strogatz network with Barab\'{a}si-Albert network. The crosses are the degree distribution of 10000 vertices and the square is the log-binning of the data. \label{fig:WSnBA}}
\end{figure}

The union of these networks exhibit real-world statistical properties like power-law-like degree distribution and high clustering coefficient \cite{loe13}. However this time we are not able to analytically determine the properties of the intersection, although simulations suggests a power-law-like degree distribution (Fig. \ref{fig:WSnBA}). Since the clustering coefficient of a subgraph is less than the graph itself, we can deduce that the intersection has low clustering coefficient given that the clustering coefficient of Barab\'{a}si-Albert network is low.

\subsubsection{Interval graphs as Substructure of Multiplex}
Since each graph in a multiplex is the intersection of interval graphs, i.e. $G^i(V,E_i) \in \mathcal{M}=I^i_1 \cap\ldots\cap I^i_d$, thus the set of overlapping edges is also the hyper-box representation of the system. Specifically the set of overlapping edges form the graph 
$H(V,E_1\cap\ldots\cap E_m) = (I^1_1 \cap \ldots \cap I^1_d) \cap \ldots \cap (I^m_1 \cap \ldots \cap I^m_{d'})$. 

Using the same idea as the previous section, we can interpret the boxicity of $H$ as the number of metrics to describe the phenomenon when information follows through all the relationships in the multiplex simultaneously. If $H$ is an interval graph, then there is a linear way for information to follow such that the conditions for all the relationships are met. 

For example we have a social multiplex with two types of relationships --- work and friends. Suppose there is a rumor regarding a company-wide action like retrenchment and there are discussions (information flow) regarding the situation. Due to the sensitivity/relevance of the issue, the discussions between colleagues are more likely to be also close friends, i.e. the rumor that spread between two people have to be connected in \emph{both} relationships. Thus the boxicity of $H$ indicates the complexity of such information flow (details in section \ref{sec:interval-propagation}).

\section{Information Propagation of Interval Graphs}
\label{sec:interval-propagation}
Information propagation is the behavior in which a property on the vertices is 
spread across the graph. In the \emph{infection model}, a vertex passes the 
property to its neighbors probabilistically at each iteration. This models the 
behavior of a virus epidemic where there is a probability for an entity to 
catch the virus from its neighbor \cite{bailey1975mathematical,
hethcote2000mathematics}.

Alternatively a vertex adopts the property under the influence of its neighbors 
when the ratio of its neighbors with the property exceeds a threshold. This is 
the \emph{influence model} and it is used to describe the nature of social 
trends like product recommendations \cite{bikhchandani1992theory,
goldenberg2001talk,granovetter1978threshold}. In general terms, vertices with 
the information (e.g. infection) are named as \emph{active} vertices, and if 
otherwise they are known as \emph{inactive} vertices.

There is a common notion with these models that information propagate along 
the edges of the network. However it is not possible in general to consider all 
the relationships in the system to map the full topology of the network. Thus 
there is a situation where information flow between non-adjacent vertices. This 
discontinuous flow of information is often assumed to be the actions of some 
confounding variables in the system and is often simulated by passing the 
information probabilistically to a random non-adjacent vertex 
\cite{myers2012information,gomez2010inferring}. 

This paper proposes the hyper-boxes representation of a graph as a 
deterministic linear framework to model the discontinuous flow of information.

\subsection{Linear Fine Structures}
\subsubsection{Overview}
The linear fine structures of a graph $G$ is the set of $m$ interval graphs 
$\{I^1(V,E_1),\ldots,I^m(V,E_m)\}$ as hyper-boxes where 
$G(V,E) = (V,E_1 \cap \ldots \cap E_m)$. The set of edges from the intervals graphs 
that are not in $G$, i.e. $E^c = (E_1 \cup \ldots \cup E_m) \setminus E$, are the 
confounding edges unobserved from the graph $G$. Thus when information 
propagate through the edges in $E^c$, it will appear from the perspective of $G$ 
that there is a discontinuous flow of information. 

In Fig. \ref{fig:interval}, the box representation of the graph on vertices $\{A,\ldots,F\}$ has boxicity 2. The box (graph) is the intersection of the bottom and left intervals, where box $A$ is enclosed by intervals $A'$ and $A''$. Suppose interval $A'$ (vertex $A$) is active and infects adjacent interval $F'$. Although $A'$ and $F'$ are adjacent, their respective boxes (vertex) are not adjacent, i.e. $A$ is not adjacent to $F$. Hence from the perspective of the graph, there is a discontinuous flow of infection between non-adjacent vertices.

For example in a marine food web, a predator feeds on species based on two 
environment niches --- the size of the prey and the depth of the ocean where 
the predator hunts. Since each of the niches can be expressed as an interval 
graph, and two predators are in competition if they feed on the same preys, 
i.e. their hyper-boxes overlaps.

Hypothetically suppose there is an increase of toxin deposits in the ocean 
and since the toxin builds up in the food chain (bioaccumulation), the toxin level of a fish is proportional to its size. Therefore the 
spread of the toxin in the ecology will appear discontinuous since the feeding 
patterns of the deep ocean marine is different from the species near the 
surface.

This framework does not obscure the context of the propagation's dynamics with 
random process, i.e. the flow of the information is well defined either by the flow 
through $E$ or $E^c$. However the trade-off is the computational 
intractability to derive the hyper-box representation from a given graph. Thus 
to demonstrate the discontinuous behavior with this linear model, random 
interval graphs are first constructed and then their intersection forms the 
observable graph $G$.

\subsubsection{Evolutionary Interval Graphs}
An evolutionary interval graph, $J_r$, parameterized by variable $r$ is to choose the 
mid-point of the intervals uniformly at random between $[0,1]$, and assign 
their length randomly from $[0,2r]$ \cite{scheinerman1990evolution}. The 
variable $r$ is also known as the ``radius" of the random mid-points. It is 
similar to the Erd\H{o}s-R\'{e}nyi graph where increasing $r$ changes the graph 
from an empty (sparse) graph to a complete (dense) graph. Similarly there 
is a phase transition for the evolutionary interval graph where the graph is 
connected with high probability. This interval graph ensemble allows us to parameterized the model such that the rate of discontinuous flow can be varied.

\subsubsection{Phase Transition}
\begin{theorem}
Let $J_r$ be an evolutionary interval graph where the intervals' length are 
chosen randomly from $[0,2r]$. If $\lim_{n\rightarrow\infty}nr/\log{n}>1$, then 
with high probability $J_r$ is connected. If otherwise $J_r$ is disconnected 
\cite{scheinerman1990evolution}.
\end{theorem}

Therefore the threshold of the phase transition is at $r \sim \log{n}/n$, and 
the following provides a more detailed structural understanding at the threshold:
\begin{theorem}\label{thm:evolution_threshold}
Let $J_r$ be an evolutionary interval graph where the intervals' length are 
chosen randomly from $[0,2r]$. If $c$ is a real constant where $r=(\log{n}+c)/n$, 
then \cite{scheinerman1990evolution}: 
\begin{equation}
Pr(J_r \mbox{ is connected}) \rightarrow e^{-e^{-c}}.
\end{equation}
\end{theorem}

To model the discontinuity of information flow, it is simpler to 
assume that the graph $G(V,E) = J^1_r \cap\ldots\cap J^m_r$ is 
connected. Since the edge set of interval graph 
$E_k \supseteq E$, thus if $G$ is connected then $J^k_r$ is connected for all 
$k$. Therefore to construct a connected graph $G$, it is necessarily (but 
insufficient) that the set of evolutionary interval graphs are connected 
(Corollary \ref{cor:hyperbox_threshold}).
\begin{corollary}\label{cor:hyperbox_threshold}
Let graph $G = J^1_r \cap\ldots\cap J^m_r$, where $J^k_r$ is the $k^{th}$ 
evolutionary interval graph and its intervals' length are chosen randomly 
from $[0,2r]$.
\begin{equation}
Pr(G \mbox{ is connected}) < Pr(J^1_r \mbox{ is connected}) \cdots Pr(J^m_r \mbox{ is connected}) \rightarrow e^{-me^{-c}}, 
\end{equation}
where $c$ is a real constant given $r=(\log{n}+c)/n$.
\end{corollary}

\begin{figure}
  \centering
  \includegraphics[width=1.0\textwidth]{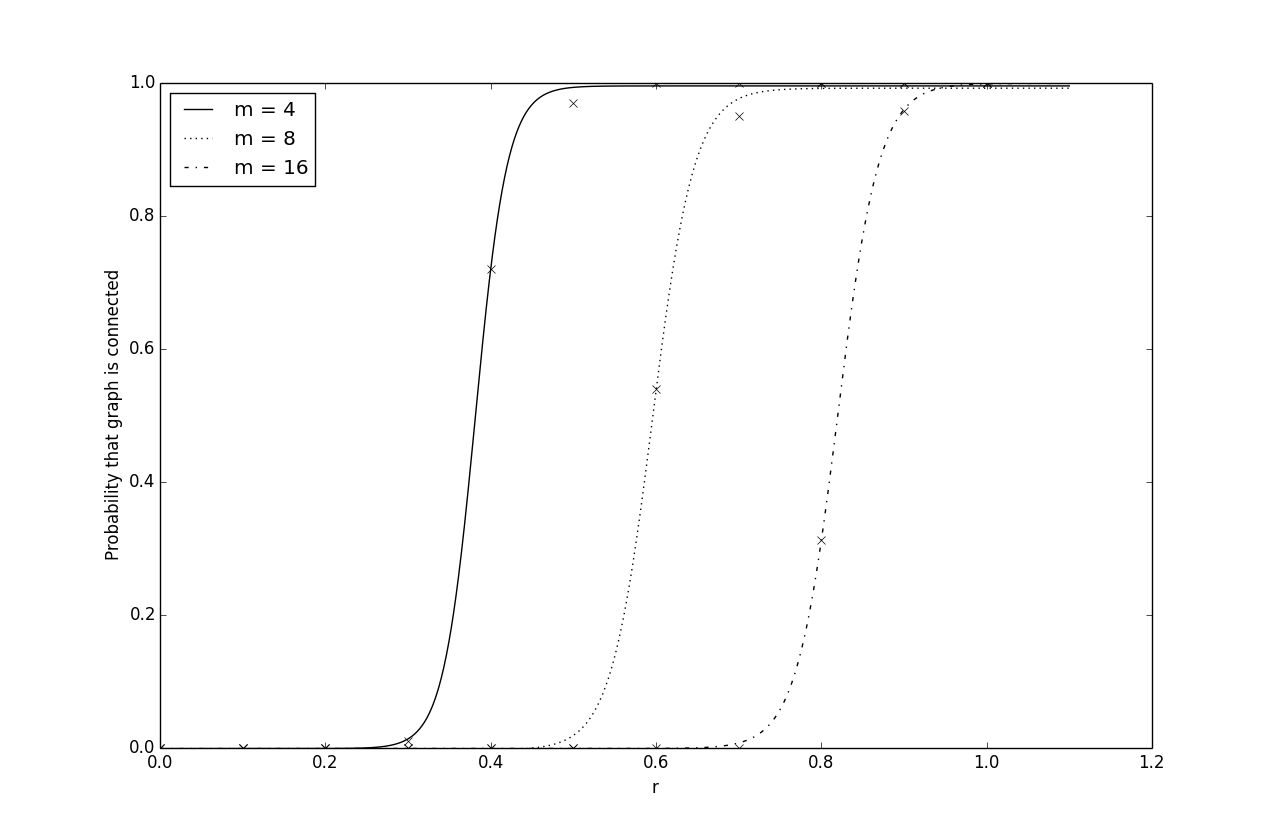}
    \caption{The probability that $G=J^1_r \cap\ldots\cap J^m_r$ is connected 
for increasing $r$. \label{fig:prob_graph_connected}}
\end{figure}

Since $\lim_{m\rightarrow\infty} e^{-me^{-c}} \rightarrow 0$, it is 
increasing harder to generate a connected graph of increasing dimension from 
a set of random evolutionary interval graphs with fixed $r$. Hence in the experiments we incrementally increase $r$ such that the graph is connected for sufficiently large $r$ 
(Fig. \ref{fig:prob_graph_connected}).

An easier alternative is to first construct a dense interval graph $J$ (of any type of ensemble like \cite{miyoshi2008scale}) and then derive the connected subgraph $G$ by randomly choosing the edges from $J$. The connectedness of $G$ on $n$ vertices can be ensured by iteratively choosing random edges, or by the critical threshold theory of Erd\H{o}s-R\'{e}nyi graph that almost all graphs with $\approx n \ln{n}$ edges are connected \cite{citeulike:4012374}.

This allows us to demonstrate this framework more efficiently without the need to repeat the process to construct random set of evolutionary interval graphs. However there is no way to parameterize this alternative such that we can vary the rate of discontinuity. Thus in the experiments we have to use the less efficient method.

\subsection{Propagation Models}
Given a connected graph $G = J^1_r \cap\ldots\cap J^m_r$, the dynamics of the 
infection model and influence model are applied to one of the interval graphs 
$J^k_r$. For example in the infection model, the active vertices in $J^k_r$ 
infects their adjacent inactive vertices with a fixed probability. Since the 
adjacent vertices in $J^k_r$ are not necessarily adjacent in $G$, the 
discontinuity of information flow can be observed from the perspective of $G$, which means that information flow is disrupted in $G$.

\subsubsection{Infection Model}
The framework of a typical infection model is the process where active vertices can transmit the infection to inactive vertices with a fixed probability per unit time. Concurrently active vertices can recover at a constant rate. The ratio between the infection rate and the recovery rate 
determines the spread of the infection (epidemic) across the network. This is also known as the SIR (susceptible-infectious-recovered) process \cite{Christley15112005}.

However in this study the rate of recovery is not relevant at this stage to understand the discontinuous flow of information (infection). This simplification is analogous to the spread of news or gossips across social networks via word of mouth \cite{sela}. The rate of infection follows the  
assumption that an inactive vertex $v$ is more susceptible to be infected if most of its neighbors are active. Thus 
\begin{equation}\label{eq:infection}
Pr(v \mbox{ will be infected})=\frac{\mbox{No. of active neighbors}}{\mbox{No. of neighbors}}.
\end{equation}

An instance of discontinuity is when a vertex is infected despite having no 
active neighbors, i.e. infected with zero probability on Eq. \ref{eq:infection}.

\subsubsection{Influence Model}
In the influence process, an inactive vertex in a network becomes active if a sufficient ratio of its adjacent vertices are active. At each time step, all inactive vertices update their status based on the number of active vertices in their neighborhood. This is similar to the behavior of fashion trends in social networks where ``non-adopters" (inactive) vertices follows the style under the 
influence of their peers.

Typically a fixed threshold $\tau$ is given in the influence model where an inactive vertex becomes active if the ratio of the number of active neighbors to neighbors is greater than $\tau$. Hence much more active vertices are required to influence a high degree vertex than a vertex with fewer neighbors. Therefore it is possible to reach an equilibrium when information no longer spread across the network, where there are insufficient active vertices to influence inactive high degree vertices \cite{kempe2003maximizing}. 

Let $v$ be one of the inactive vertex in a network with threshold $\tau_v$ and the set of its neighbors be $N_v$. In the generalized model, each neighbor $u \in N$ has a weighted influence $w_{u,v}$ on vertex $v$, such that $\sum_{u\in N_v}w_{u,v} = 1$. Hence in each iteration $v$ will be active if 
\begin{equation}\label{eq:influence}
\sum_{u\in N'_v}w_{u,v} \geq \tau_v,
\end{equation}
where $N'_v$ is the set of active neighbors of $v$.

In the experiments, the following simplifications are assumed. The thresholds for every vertex are equal, i.e. $\tau_1=\ldots=\tau_n=\tau=0.5$, and the weighted influence is balance, 
i.e. $w_{u,v}=1/|N_v|$. From the perspective of the graph $G$, if an inactive vertex becomes active despite not fulfilling Eq. \ref{eq:influence}, then this phenomenon is defined as an instance of discontinuity in the information flow. In this model, discontinuity is also defined if $v$ remains inactive even when it is above the threshold.

\subsection{Experimental Results}
\begin{figure}
  \centering
  \includegraphics[width=1.0\textwidth]{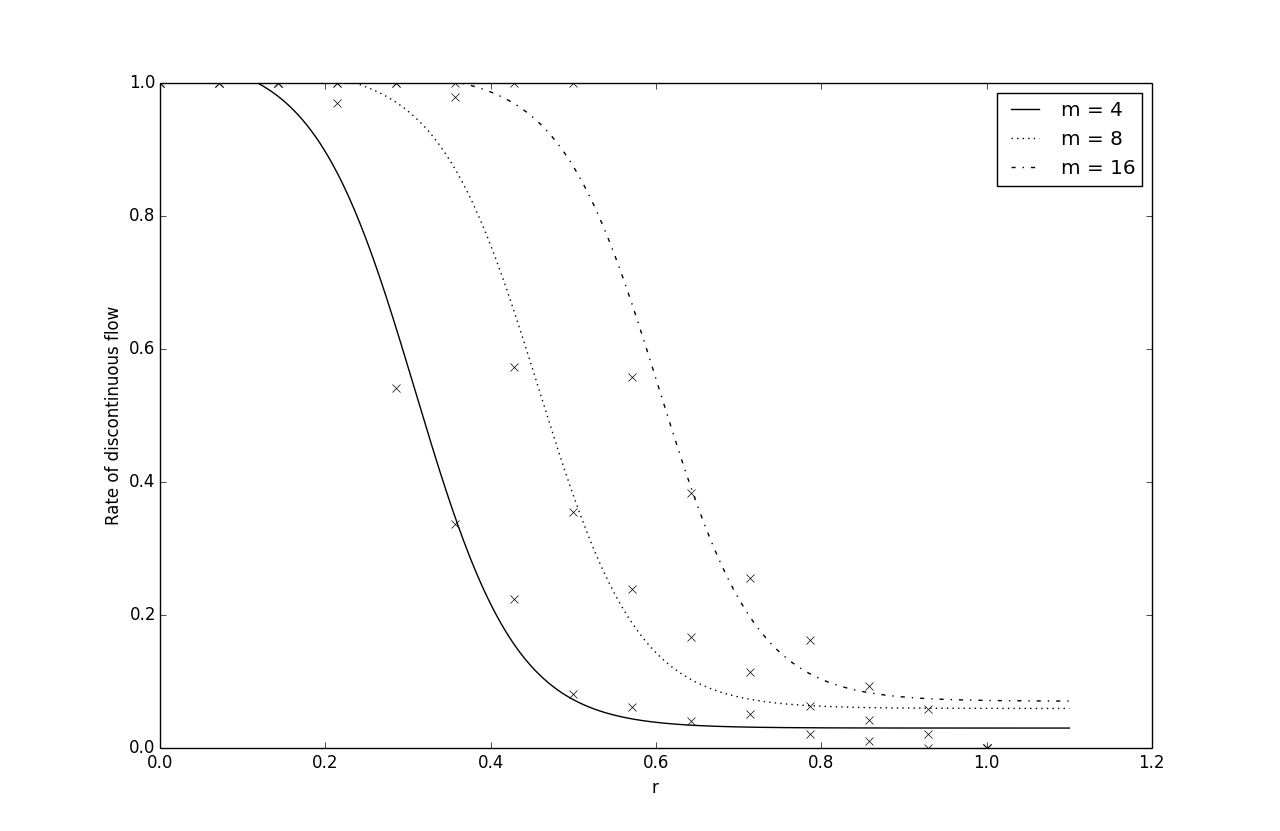}
  \includegraphics[width=1.0\textwidth]{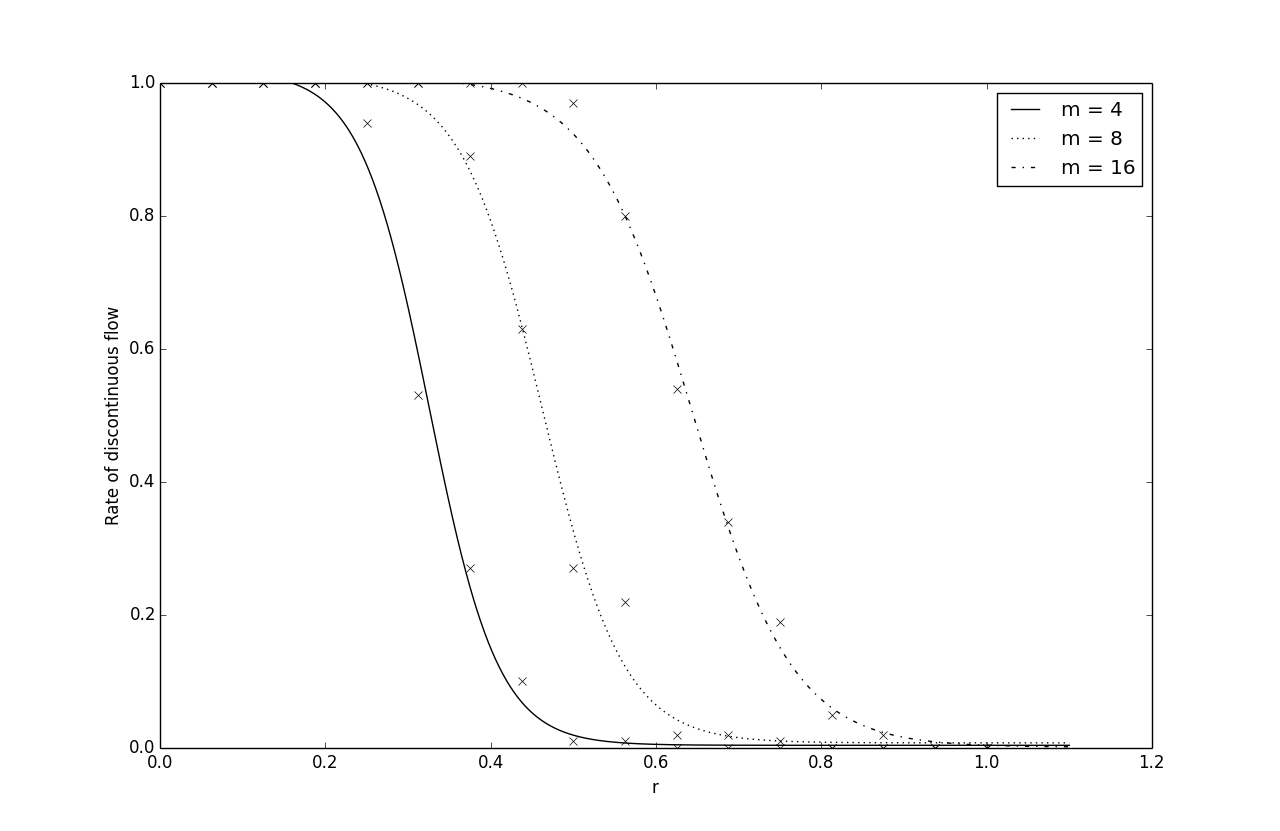}
  \caption{The rate of discontinuity observed in $G=J^1_r \cap\ldots\cap J^m_r$ 
when the infection (top plot) dynamics is applied to a random interval graph $J^k_r$. Similarly the bottom plot shows the rate of discontinuity when the influence dynamics is applied to a random evolutionary interval graph. 
\label{fig:discontinuity}}
\end{figure}

Fig. \ref{fig:discontinuity} is a proof of concept that it is possible to simulate any rate of discontinuity with evolutionary interval graphs. We define the rate of discontinuity $=1$ when the graph is disconnected so that the plot fits a Sigmoid-like function. Moreover if we assume that all the vertices eventually have to be active, then isolated vertices that becomes active must be under the influence of some discontinuity. Regardless it is more meaningful (at this point) to look at the propagation when $r$ is large.

As $r$ increases the graph (from the intersection of all the interval graphs) becomes denser and every vertex has an edge connecting to most of the other vertices. Thus the effects of discontinuity is not apparent. For example under the infection model, a vertex tends to have more neighbors in the interval graph than the graph itself. This affects Eq. \ref{eq:infection}, i.e. the probability that a vertex will be infected from the perspective of the graph is different from the ``true" measurement from the perspective of the interval graph.

Although in real-world system, the interval graphs do not necessarily belong to the ensemble of evolutionary interval graphs. The experiments is simply to support the hypothesis that the framework of interval graphs is a deterministic way to model discontinuity in network propagation. However Fig. \ref{fig:discontinuity} also allows us to suggest that the greater the boxicity (complexity), the more likely we expect the network to exhibit discontinuity. 

The explanation is that it is unlikely (by random chance) the edges in one of the interval graphs to be in all the other interval graphs. Hence the greater the boxicity, the more likely an edge will not be reflected on the graph itself. Therefore there will be many edges from all the interval graphs that are not on the graph and increase the change of discontinuity.

However most importantly it is also possible that the propagation ``switch" dimension, i.e. instead of continuing to follow the edges of a particular interval graph, the information can spread via the edges of another interval graph. For example in Fig. \ref{fig:interval}, information can flow from the horizontal interval $A'$ to $B'$ and then ``jump" to $E''$ via the vertical interval. Thus the greater the boxicity, the more dimensions (degree of freedom) for the propagation to switch about and further increase the rate of discontinuity.

\subsection{Discussion}
Although in principle interval graphs can be used as a framework to simulate the discontinuity of information propagation, it is important to figure out its role in our existing understanding of Network Science. For example what insights it can deliver that existing models fail to, vice versa. More importantly even if a model fits the data, it must have meaningful contexts to relate to the dynamics of the system. 

One of the advantages of a deterministic model is that the simulations can easily be repeatable once the direction of the flow is chosen. This property can be mimic easily by other probabilistic models by choosing a pseudo-random number generator with a fixed seed. However the slight difference is that in our deterministic framework, the dynamics of Eq. \ref{eq:infection} and Eq. \ref{eq:influence} can remain probabilistic. This implies this model (as oppose to others) can control the general direction of the information flow, but not the propagation dynamics.

We emphasized that this framework is an \emph{alternative} and not a \emph{replacement} for existing models as we understand the subjectivity of embedding relational information in networks. Given that there are meaningful contexts of intervals graphs with complex systems like Ecology and Bioinformatics, we believe that there should be valid applications in the broader scope of Network Science such that this framework aptly models the system. For example time dependent systems like the EEG or fMRI time series of brain networks.

\section{``Approximating" Boxicity Using Communities Detection}
In the previous sections, this paper presents the alternative perspectives of complexity using interval graphs for Network Science. However the applicability for real world systems will remain limited since to determine a network's boxicity is a NP-complete problem.

Although a network's boxicity can be bounded by their maximum degree (section \ref{sec:hyperbox}), the bound is generally not tight. Moreover the boxicity of a network is an unstable and non-monotonous function where it fluctuates unpredictability when new edges/nodes are added to the network. Hence the intolerance to experimental errors further challenge the applicability of boxicity.

\subsection{Minimum Boxicity of Network from its Communities}
We propose that communities detection is a key strategy to resolve the above problems. It is similar to optimizing the Hamiltonian Walk problem by simplifying a network into modular structures 
\cite{Jones}. Firstly the boxicity of a community (induced subgraph) is a simpler problem since it is a smaller network. This can be more meaningful for problems where the understanding of the individual communities is more important than the entire network.

Since the boxicity of a graph is at least the boxicity of its subgraph \cite{roberts69}, thus: 
\begin{lemma}\label{lemma:sub_boxicity}
$Boxicity(G) \geq max_{g \in C} Boxicity(g)$, where $C$ is the set of communities in graph $G$.
\end{lemma}

\begin{figure}
  \centering
  \includegraphics[width=1.1\textwidth]{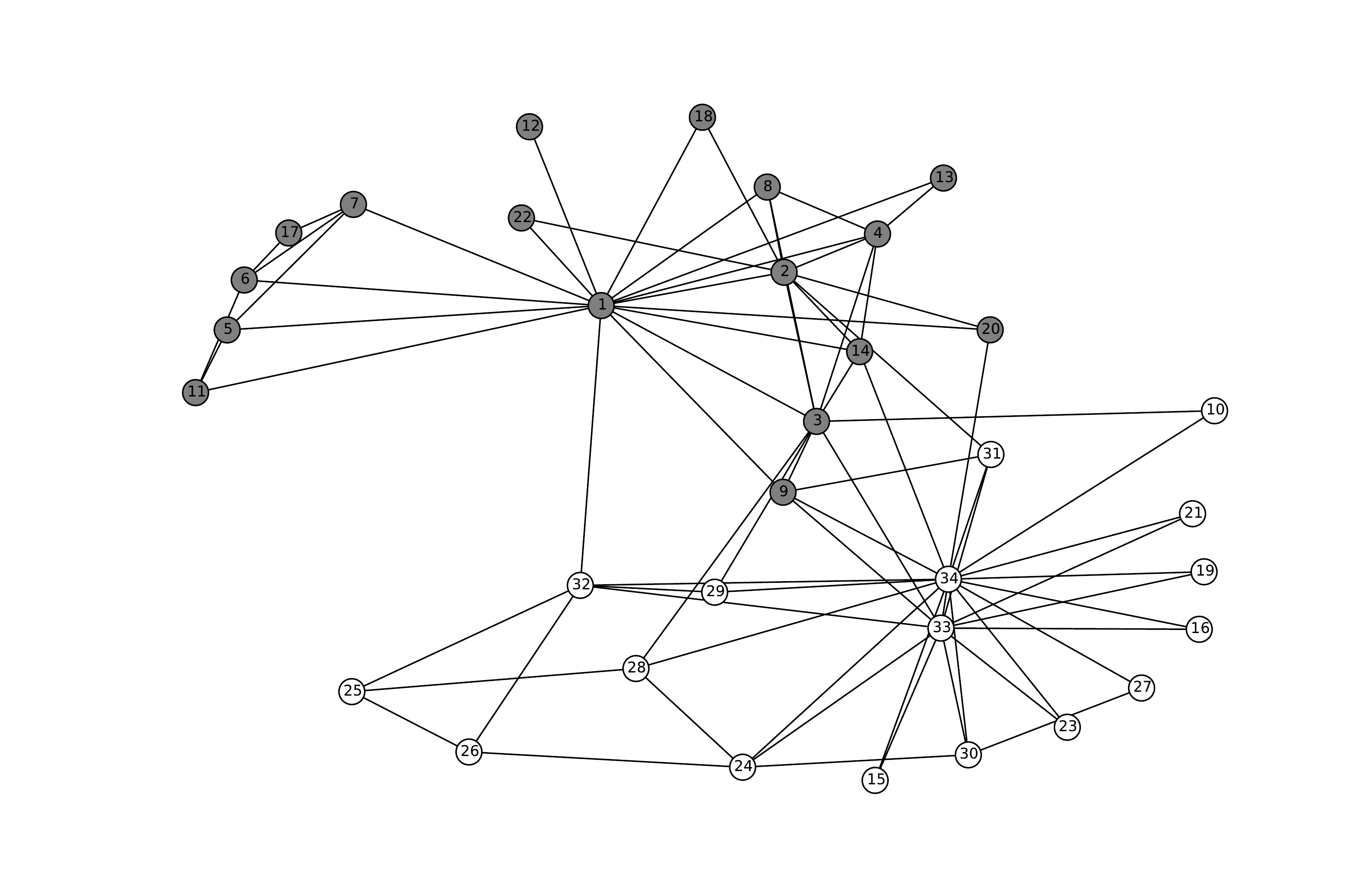}
  \includegraphics[width=1.1\textwidth]{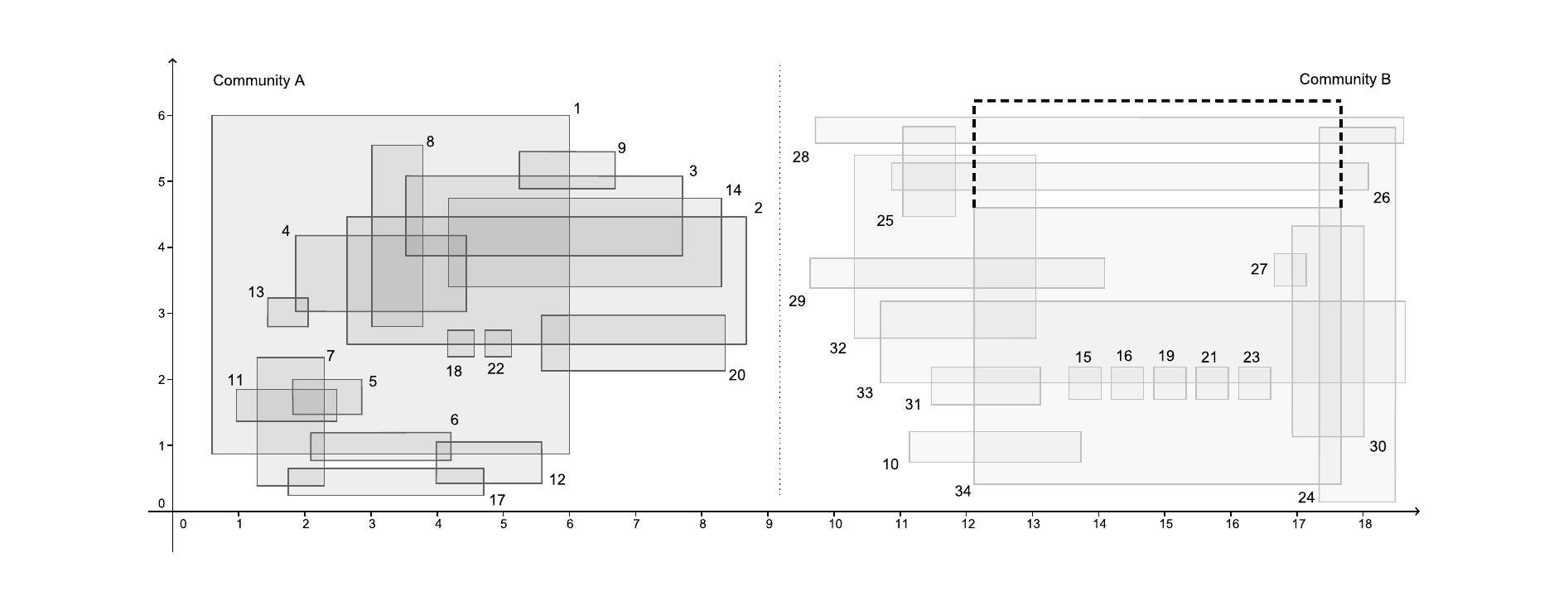}
  \caption{(Top) The Zachary Karate Club Network where communities $A$ and $B$ are the shaded and non-shaded nodes respectively. (Bottom) The hyperbox representation of community $A$ and $B$. The dashed line represents Community $B$ in 2-dimension (boxicity $=2$) if vertex 34 is adjacent to vertex 26. Since it is not, hence we need the third dimension such that the box 34 can overlap box 28 while ``bridging over" (bypass) box 26. 

The boxes are aligned in a way such that vertices that connects to the other communities are near the center. For example the vertices in community $A$ have to route via vertices $\{1,2,3,9,14,20\}$ to get to community $B$. Similarly the vertices in community $B$ have to route via vertices $\{10,28,29,31,32,33,34\}$ to reach community $A$. Since we divide the network into two smaller communities, this constrain is more sensible when we try to ``join" the communities' hyperboxes.
\label{fig:zachary}}
\end{figure}

For instance there are two communities in the Zachary Karate Club Network with 17 vertices in community $A$ and 16 vertices in community $B$ (Top diagram in Fig. \ref{fig:zachary}). The network is not an interval graph as vertices $\{24,25,26,28\}$ is not chordal (theorem \ref{interval_test}). Now that the communities are small, we are able to easily deduce that community $A$ has boxicity $=2$ and community $B$ has boxicity $>2$ (no solution found via exhaustive search). 

But given that community $B$ is a planar graph, its boxicity $\leq 3$ \cite{thomassen1986}. This implies the boxicity of community $B$ is 3. Therefore the boxicity of the Zachary Karate Club Network $\geq 3$ (lemma \ref{lemma:sub_boxicity}). The bottom diagram in Fig. \ref{fig:zachary} shows one of the possible hyperbox representations of the communities. Since we have to eventually combine these partial solutions, it will be useful to constrain the partial solutions where the vertices of a community that connects to the other communities have to be aligned along the boundaries of the hyperboxes.

\begin{sidewaysfigure}
\centering
\includegraphics{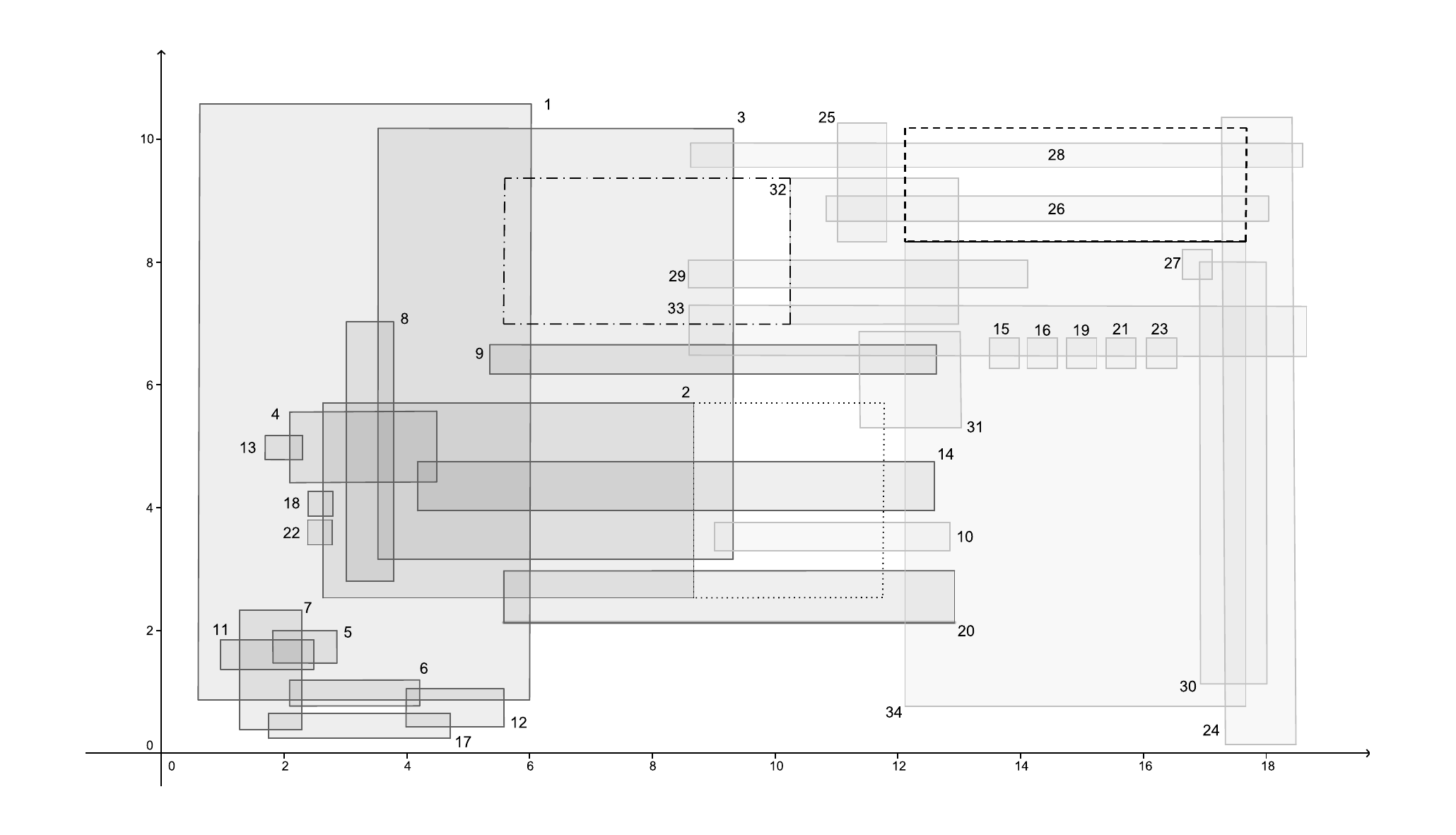}
\caption{The hyperbox representation of the Zachary Karate Club Network. The boxes have to be transformed such that the connection between communities $A$ and $B$ is complete. The dashed line refers to the extension of box 34 to overlap box 28 such that it bypass box 26. The dotted line refers to the extension of box 2 to overlap box 31 such that it bypass box 10. The dot-dashed line refers to the extension of box 32 to overlap box 1 such that it bypass box 3. \label{fig:zachary3d}}
\end{sidewaysfigure}

Fig. \ref{fig:zachary3d} slightly rearranged the hyperboxes in Fig. \ref{fig:zachary} such that the boxes at the boundaries the communities can be easily combined. From the figure, we can conclude that it does not take more than $3$-dimensions to combine the communities. Hence the boxicity of Zachary Karate Club Network is 3.

\subsection{Boxcity of the Communities' Interaction Network}
Lastly communities detection allows us to look at the problem from a more general perspective. It represents a broad overview on how information flows from one community to another. The complexity (boxicity) of this modular structure is important to understand the information propagation of a network.

This is done by coarsening the network $G$ with a new network $H$ where the vertices in $H$ represents the communities of $G$, and the vertices in $H$ are adjacent if and only if their corresponding communities in $G$ are connected. Since $H$ is a smaller network than $G$, the computation of the boxicity of $H$ is easier. This process can be repeated on $H$ until we get the desired granularity.

In our previous example, the Zachary Karate Club Network, there is only two communities and they are connected. Hence the coarse network is just a complete graph on two vertices, i.e. an interval graph with boxicity $=1$. This implies that the information flow has low complexity where there is a linear flow from one community to another. 

\subsection{Boxicity with Experimental Noise}
The conclusion from the previous example is trivial since there are only two communities. However it is interesting to note that the conclusion remains the same when we remove/add (a small number of) edges from/to the network. These modifications can represent the noise in the experiments and hence more relevant for scientific applications. 

The instability of boxicity due to noise was the reason to why \emph{Quasi-Interval Graph}, $Q$,  was introduced. It is a graph with boxicity $>1$ that can be expressed as an interval graph by adding or subtracting some edges as experimental errors from $Q$. It is useful for systems where there are strong qualitative evidences that they have linear structure  \cite{stouffer2006robust,cattin2004phylogenetic,mouillot2008high}. This can be done by finding the minimum  number of edges to 1) add to $Q$ \cite{kaplan1999bounded,
kaplan1999tractability,natanzon2000polynomial}, 2) remove from $Q$ \cite{goldberg1995four} or 3) a mixture of both types of errors \cite{lu2003clustering}. For example community $B$ in the Zachary Karate Club Network will have boxicity $=2$ (instead of 3) if vertices 26 and 34 are adjacent (Fig. \ref{fig:zachary}). 

However it is still a hard problem to minimize the number of modification over the entire network such that its boxicity is also minimized. Thus is more intuitive and easier to understand the general dynamics of a system with the coarsen network than the precise boxicity of the network.

\section{Discussions}
Information flow across interval graphs is an established model in ecology research to understand the stability of complex systems. However to relate this to the broader applications in Network Science has yet to be attempted. For instance the simulations show that as a proof of concept, interval graphs are viable linear fine structures to model the real world characteristic of discontinuous information propagation. The advantage of this framework is that the intuitions of the dynamics are not obscure by random processes.

In addition we show that the methodologies in Network Science can be useful to resolve some of the challenges of interval graph and boxicity. For example communities detection algorithms  modularize the network such that the problem of boxicity can be simplified. Furthermore the communities allow us to focus on the complexity of the general network topology rather than the details within the communities which are prone to experimental errors.

Given the growing interests for multi-relational networks, we believe that interval graphs will appeal to scientists and complement their research. In addition the modern methodologies and tools from Network Science can further improve the computational hardness of boxicity. Thus revisiting interval graphs will broaden our perspective of complexity theory.

\bibliographystyle{tCPH}
\bibliography{database}

\begin{thebibliography}{10}
\newcommand{\noopsort}[1]{}
\newcommand{\printfirst}[2]{#1}
\newcommand{\singleletter}[1]{#1}
\newcommand{\switchargs}[2]{#2#1}
\providecommand{\url}[1]{\normalfont{#1}}
\providecommand{\urlprefix}{Available at }

\bibitem{benzer1959topology}
S. Benzer, \emph{On the topology of the genetic fine structure}, Proceedings of
  the National Academy of Sciences of the United States of America 45 (1959),
  p. 1607.

\bibitem{opac-b1129706}
J.E. Cohen, \emph{Food webs and niche space}, Monographs in population biology,
  Princeton Univ. Press, Princeton NJ, 1978, includes index.

\bibitem{eklof2013dimensionality}
A. Ekl{\"o}f, U. Jacob, J. Kopp, J. Bosch, R. Castro-Urgal, N.P. Chacoff, B.
  Dalsgaard, C. Sassi, M. Galetti, P.R. Guimar{\~a}es, \emph{et~al.}, \emph{The
  dimensionality of ecological networks}, Ecology letters 16 (2013), pp.
  577--583.

\bibitem{jordan2004network}
F. Jord{\'a}n and I. Scheuring, \emph{Network ecology: topological constraints
  on ecosystem dynamics}, Physics of Life Reviews 1 (2004), pp. 139--172.

\bibitem{carlisle1995k}
M.C. Carlisle and E.L. Lloyd, \emph{On the k-coloring of intervals}, Discrete
  Applied Mathematics 59 (1995), pp. 225--235.

\bibitem{kolen2007interval}
A.W. Kolen, J.K. Lenstra, C.H. Papadimitriou, and F.C. Spieksma, \emph{Interval
  scheduling: A survey}, Naval Research Logistics (NRL) 54 (2007), pp.
  530--543.

\bibitem{gattass1981}
E.A. Gattass and G.L. Nemhauser, \emph{An application of vertex packing to data
  analysis in the evaluation of pavement deterioration}, Operations Research
  Letters 1 (1981), pp. 13--17.

\bibitem{jungck1982computer}
J.R. Jungck, G. Dick, and A.G. Dick, \emph{Computer-assisted sequencing,
  interval graphs, and molecular evolution}, Biosystems 15 (1982), pp.
  259--273.

\bibitem{biedl2004finding}
T. Biedl, B. Brejov{\'a}, E.D. Demaine, A.M. Hamel, A. L{\'o}pez-Ortiz, and T.
  Vina{\v{r}}, \emph{Finding hidden independent sets in interval graphs},
  Theoretical Computer Science 310 (2004), pp. 287--307.

\bibitem{Kivela2013}
M. Kivel\"{a}, A. Arenas, M. Barthelemy, J.P. Gleeson, Y. Moreno, and M.A.
  Porter, \emph{Multilayer networks}, Journal of Complex Networks  (2014).

\bibitem{boccaletti2014structure}
S. Boccaletti, G. Bianconi, R. Criado, C. del  Genio, J.
  G{\'o}mez-Garde{\~n}es, M. Romance, I. Sendi{\~n}a-Nadal, Z. Wang, and M.
  Zanin, \emph{The structure and dynamics of multilayer networks}, Physics
  Reports  (2014).

\bibitem{terry1999topics}
T.A. McKee and F.R. McMorris, \emph{Topics in Intersection Graph Theory},
  Monographs on Discrete Mathematics and Applications, Society for Industrial
  and Applied Mathematics, 1999.

\bibitem{fishburn1985interval}
P.C. Fishburn, \emph{Interval graphs and interval orders}, Discrete mathematics
  55 (1985), pp. 135--149.

\bibitem{Lekkeikerker1962}
B.J. Lekkeikerker C., \emph{Representation of a finite graph by a set of
  intervals on the real line}, Fundamenta Mathematicae 51 (1962), pp. 45--64.

\bibitem{chandran2010geometric}
L.S. Chandran, M.C. Francis, and N. Sivadasan, \emph{Geometric representation
  of graphs in low dimension using axis parallel boxes}, Algorithmica 56
  (2010), pp. 129--140.

\bibitem{stouffer2006robust}
D.B. Stouffer, J. Camacho, and L.A.N. Amaral, \emph{A robust measure of food
  web intervality}, Proceedings of the National Academy of Sciences 103 (2006),
  pp. 19015--19020.

\bibitem{adiga2008lower}
A. Adiga, L. Chandran, and N. Sivadasan, \emph{Lower bounds for boxicity},
  Combinatorica  (2014), pp. 1--25.

\bibitem{sunil2008boxicity}
L. Sunil~Chandran, M.C. Francis, and N. Sivadasan, \emph{Boxicity and maximum
  degree}, Journal of Combinatorial Theory, Series B 98 (2008), pp. 443--445.

\bibitem{Freeman83}
L.C. Freeman, \emph{Spheres, cubes and boxes: Graph dimensionality and network
  structure}, Social Networks 5 (1983), pp. 139 -- 156.

\bibitem{zachary1977ifm}
W.W. Zachary, \emph{{An information flow model for conflict and fission in
  small groups}}, Journal of Anthropological Research 33 (1977), pp. 452--473.

\bibitem{loe13}
C.W. Loe and H.J. Jensen, \emph{Edge union of networks on the same vertex set},
  Journal of Physics A: Mathematical and Theoretical 46 (2013), p. 245002.

\bibitem{Bianconi13}
G. Bianconi, \emph{Statistical mechanics of multiplex networks: Entropy and
  overlap}, Phys. Rev. E 87 (2013), p. 062806.

\bibitem{chen2012degree}
A.L. Chen, F.J. Zhang, and H. Li, \emph{The degree distribution of the random
  multigraphs}, Acta Mathematica Sinica, English Series 28 (2012), pp.
  941--956.

\bibitem{Lee12}
K.M. Lee, J.Y. Kim, W. kuk  Cho, K.I. Goh, and I.M. Kim, \emph{Correlated
  multiplexity and connectivity of multiplex random networks}, New Journal of
  Physics 14 (2012), p. 033027.

\bibitem{RanolaASSL10}
J.M.O. Ranola, S. Ahn, M.E. Sehl, D.J. Smith, and K. Lange, \emph{A poisson
  model for random multigraphs.}, Bioinformatics 26 (2010), pp. 2004--2011.

\bibitem{barabasia99}
A.L. Barabasi and R. Albert, \emph{Emergence of scaling in random networks},
  Science 286 (1999), pp. 509--512.

\bibitem{dorogovtsev2000structure}
S.N. Dorogovtsev, J.F.F. Mendes, and A.N. Samukhin, \emph{Structure of growing
  networks with preferential linking}, Physical Review Letters 85 (2000), p.
  4633.

\bibitem{Pennock02winnersdont}
D.M. Pennock, G.W. Flake, S. Lawrence, E.J. Glover, and C.L. Giles,
  \emph{Winners don't take all: Characterizing the competition for links on the
  web}, Proceedings of the national academy of sciences 99 (2002), pp.
  5207--5211.

\bibitem{Stumpf05}
M.P.H. Stumpf, C. Wiuf, and R.M. May, \emph{Subnets of scale-free networks are
  not scale-free: Sampling properties of networks}, Proceedings of the National
  Academy of Sciences of the United States of America 102 (2005), pp.
  4221--4224.

\bibitem{bailey1975mathematical}
N.T. Bailey, \emph{et~al.}, \emph{The mathematical theory of infectious
  diseases and its applications}, Charles Griffin \& Company Ltd, 5a Crendon
  Street, High Wycombe, Bucks HP13 6LE., 1975.

\bibitem{hethcote2000mathematics}
H.W. Hethcote, \emph{The mathematics of infectious diseases}, SIAM review 42
  (2000), pp. 599--653.

\bibitem{bikhchandani1992theory}
S. Bikhchandani, D. Hirshleifer, and I. Welch, \emph{A theory of fads, fashion,
  custom, and cultural change as informational cascades}, Journal of political
  Economy  (1992), pp. 992--1026.

\bibitem{goldenberg2001talk}
J. Goldenberg, B. Libai, and E. Muller, \emph{Talk of the network: A complex
  systems look at the underlying process of word-of-mouth}, Marketing letters
  12 (2001), pp. 211--223.

\bibitem{granovetter1978threshold}
M. Granovetter, \emph{Threshold models of collective behavior}, American
  journal of sociology  (1978), pp. 1420--1443.

\bibitem{myers2012information}
S.A. Myers, C. Zhu, and J. Leskovec, \emph{Information diffusion and external
  influence in networks}, in \emph{Proceedings of the 18th ACM SIGKDD
  international conference on Knowledge discovery and data mining}, 2012, pp.
  33--41.

\bibitem{gomez2010inferring}
M. Gomez~Rodriguez, J. Leskovec, and A. Krause, \emph{Inferring networks of
  diffusion and influence}, in \emph{Proceedings of the 16th ACM SIGKDD
  international conference on Knowledge discovery and data mining}, 2010, pp.
  1019--1028.

\bibitem{scheinerman1990evolution}
E.R. Scheinerman, \emph{An evolution of interval graphs}, Discrete Mathematics
  82 (1990), pp. 287--302.

\bibitem{miyoshi2008scale}
N. Miyoshi, T. Shigezumi, R. Uehara, and O. Watanabe, \emph{Scale free interval
  graphs}, in \emph{Algorithmic Aspects in Information and Management},
  Springer,  2008, pp. 292--303.

\bibitem{citeulike:4012374}
P. Erd\H{o}s and A. R\'{e}nyi, \emph{{On random graphs, I}}, Publicationes
  Mathematicae (Debrecen) 6 (1959), pp. 290--297,
  \urlprefix\url{http://www.renyi.hu/\~{}p\_erdos/Erdos.html\#1959-11}.

\bibitem{Christley15112005}
R.M. Christley, G.L. Pinchbeck, R.G. Bowers, D. Clancy, N.P. French, R.
  Bennett, and J. Turner, \emph{Infection in social networks: Using network
  analysis to identify high-risk individuals}, American Journal of Epidemiology
  162 (2005), pp. 1024--1031.

\bibitem{sela}
O.H. Sela~A. and B.G. I., \emph{Information Spread in a Connected World}, in
  \emph{Proceedings of Collective Intelligence 2014}, MIT Boston, 2014.

\bibitem{kempe2003maximizing}
D. Kempe, J. Kleinberg, and {\'E}. Tardos, \emph{Maximizing the spread of
  influence through a social network}, in \emph{Proceedings of the ninth ACM
  SIGKDD international conference on Knowledge discovery and data mining},
  2003, pp. 137--146.

\bibitem{Jones}
B.M. Bui-Xuan and N.S. Jones, \emph{How modular structure can simplify tasks on
  networks: parameterizing graph optimization by fast local community
  detection}, Proceedings of the Royal Society of London A: Mathematical,
  Physical and Engineering Sciences 470 (2014).

\bibitem{roberts69}
F.S. Roberts, \emph{On the boxicity and cubicity of a graph}, Recent Progress
  in Combinatorics(W. T. Tutte, ed.)  (1996), pp. 301--310.

\bibitem{thomassen1986}
C. Thomassen, \emph{Interval representations of planar graphs}, Journal of
  Combinatorial Theory, Series B 40 (1986), pp. 9--20.

\bibitem{cattin2004phylogenetic}
M.F. Cattin, L.F. Bersier, C. Bana{\v{s}}ek-Richter, R. Baltensperger, and J.P.
  Gabriel, \emph{Phylogenetic constraints and adaptation explain food-web
  structure}, Nature 427 (2004), pp. 835--839.

\bibitem{mouillot2008high}
D. Mouillot, B.R. Krasnov, and R. Poulin, \emph{High intervality explained by
  phylogenetic constraints in host-parasite webs}, Ecology 89 (2008), pp.
  2043--2051.

\bibitem{kaplan1999bounded}
H. Kaplan and R. Shamir, \emph{Bounded degree interval sandwich problems},
  Algorithmica 24 (1999), pp. 96--104.

\bibitem{kaplan1999tractability}
H. Kaplan, R. Shamir, and R.E. Tarjan, \emph{Tractability of parameterized
  completion problems on chordal, strongly chordal, and proper interval
  graphs}, SIAM Journal on Computing 28 (1999), pp. 1906--1922.

\bibitem{natanzon2000polynomial}
A. Natanzon, R. Shamir, and R. Sharan, \emph{A polynomial approximation
  algorithm for the minimum fill-in problem}, SIAM Journal on Computing 30
  (2000), pp. 1067--1079.

\bibitem{goldberg1995four}
P.W. Goldberg, M.C. Golumbic, H. Kaplan, and R. Shamir, \emph{Four strikes
  against physical mapping of dna}, Journal of Computational Biology 2 (1995),
  pp. 139--152.

\bibitem{lu2003clustering}
W.F. Lu and W.L. Hsu, \emph{A clustering algorithm for interval graph test on
  noisy data}, in \emph{Experimental and Efficient Algorithms}, Springer,
  2003, pp. 195--208.

\end{thebibliography}

\end{document}